\documentclass[
 reprint,
 aps, showkeys, showpacs
]{revtex4-1}
\usepackage{graphicx}
\usepackage{dcolumn}
\usepackage{bm}
\usepackage{siunitx}
\usepackage[fleqn]{amsmath}
\usepackage{amssymb}
\usepackage[utf8]{inputenc}
\usepackage{epstopdf}
\usepackage{float}
\begin{document}
\preprint{APS/123-QED}

\title{Particle transport across a channel via an oscillating potential}

\author{Yizhou Tan}
\author{Jannes Gladrow}
\author{Ulrich F. Keyser}
\affiliation{
 Cavendish laboratory, University of Cambridge, Cambridge CB3 0HE, United Kingdom
}

\author{Leonardo Dagdug}
\affiliation{
 Departamento de Fisica, Universidad Autonoma Metropolitana-Iztapalapa, 09340 Mexico City, Mexico
}

\author{Stefano Pagliara}
\email{s.pagliara@exeter.ac.uk}
\affiliation{
Living Systems Institute, University of Exeter, Exeter EX4 4QD, United Kingdom
}
\affiliation{
 Cavendish laboratory, University of Cambridge, Cambridge CB3 0HE, United Kingdom
}

\date{\today}

\begin{abstract}

Membrane protein transporters alternate their substrate-binding sites between the extracellular and cytosolic side of the membrane according to the alternating access mechanism. Inspired by this intriguing mechanism devised by nature, we study particle transport through a channel coupled with an energy well that oscillates its position between the two entrances of the channel.  We optimize particle transport across the channel by adjusting the oscillation frequency. At the optimal oscillation frequency, the translocation rate through the channel is a hundred times higher with respect to free diffusion across the channel. Our findings reveal the effect of time dependent potentials on particle transport across a channel and will be relevant for membrane transport and microfluidics application.

\end{abstract}

\keywords{}

\maketitle

\section{Introduction}
Transport proteins are ubiquitously expressed in all kingdoms of life and allow for the continuous exchange of ions and nutrients across cell membranes~\cite{alberts1995molecular}. A feature common to all transporters is their capability to bind their substrate. The number, position and strength of the substrate-binding sites can be optimized in order to maximize substrate exchange across the cell membrane~\cite{Kasianowicz2006a}. The physical mechanisms underlying transport optimization have been extensively investigated experimentally ~\cite{Benz1986,Ward1992,Meyer1997,Kullman2002a,Pagliara2014e,Pagliara2013,Horner2015}, by molecular dynamics simulations~\cite{Jensen2002}, and independently rationalized by a continuum diffusion model based on the
Smoluchowski equation~\cite{Berezhkovskii2005}, a discrete stochastic model~\cite{Kolomeisky2007} and a general kinetic model~\cite{Zilman2009a}. 

However, these studies do not take into account a fundamental hallmark shared by several transporters that is their capability to alternate their substrate-binding sites between the extracellular and cytosolic side of the membrane according to the alternating access mechanism proposed fifty years ago~\cite{Jardetzky1966}.  A simplified alternated particle transport mechanism can be achieved by modulating the energy landscape in which particles diffuse. Indeed, optical potentials modulated in time have been employed to study particle diffusion~\cite{Lee2006,Juniper2016}, to induce thermal ratchets~\cite{Faucheux1995,Leea} to direct ~\cite{Bleil2007} and sort Brownian particles~\cite{Gorre-Talini1997,MacDonald2003,Jonas2008,Xiao,Ladavac}, to study particle escape and synchronization~\cite{Simon1992}, to investigate stochastic resonance and resonant activation~\cite{Babic2004a,Schmitt2006}. However, to the best of our knowledge, the effect of oscillating potentials on particle transport across one-dimensional (1D) channels remains to be investigated.

In this letter, inspired by the naturally occurring alternating access mechanism, we use our previously introduced experimental model system~\cite{Pagliara2013,Pagliara2014e,Misiunas2015} to couple a modulated potential in a quasi 1D microfluidic channel. Specifically, we use holographic optical tweezers (HOTs)~\cite{Padgett2011b,Bowman2014} to create an optical potential that oscillates in time between the two entrances of the channel. We find that (i) there is an optimal oscillation frequency that maximizes the particle transport rate through the channel; (ii) at this oscillation frequency, the particle transport rate is two orders of magnitude larger with respect to free diffusion; (iii) channel occupancy increases with oscillation frequency and (iv) the optimal oscillation frequency is the one that resonates with particle diffusion across the region between the centres of the energy well positions. 

\begin{figure}
	\includegraphics[width=\linewidth]{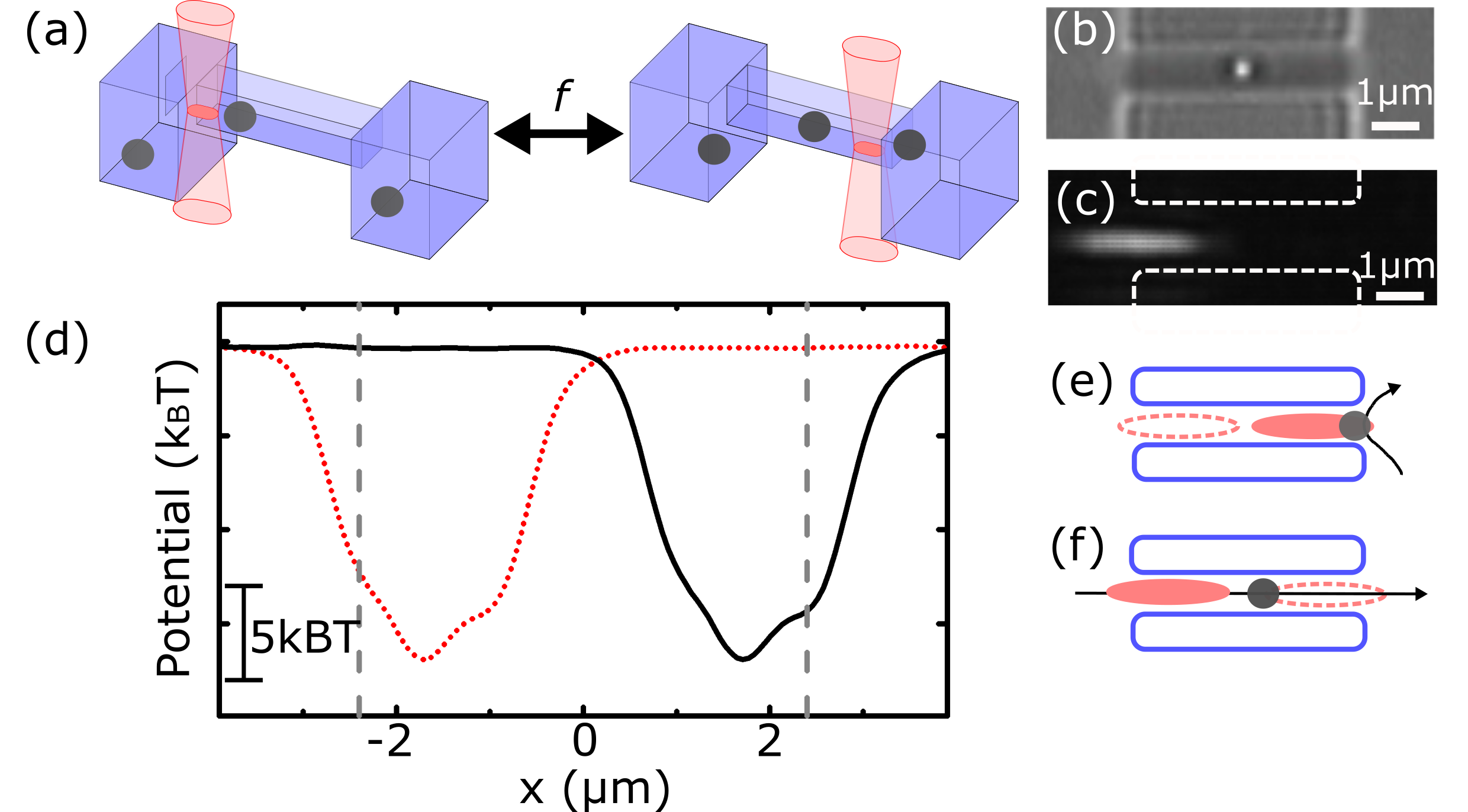}
	\caption{\label{Fig1} (a) Schematics illustrating the oscillation of the position of a laser line trap between the left and right hand side of a channel at a frequency \textit{f}. (b) Bright-field image of a 510 nm polystyrene particle diffusing in a microfluidic channel. (c) Corresponding dark-field image showing the intensity distribution of a laser line trap positioned at the left entrance of the microfluidic channel. The dashed lines highlight the channel contours. (d) Oscillating energy potential created when the laser trap is positioned at the left (dotted line) and at the right channel entrance (solid line). The potential extension, depth and shape are estimated from the intensity distribution of the laser traps. The vertical lines indicate the channel entrances. (e) Schematics illustrating particle return (e) and translocation (f) events.}
\end{figure}

\section{Experimental methods}
Our microfluidic devices are fabricated as previously reported~\cite{Pagliara2011,Pagliara2014d}. They consist of two 3D reservoirs with a depth of 12 $\mu$m separated by a polydimethylsiloxane barrier and connected by an array of microfluidic channels. Each channel has a cross section of around 0.9$\times$0.9 $\mu$m$^2$ and a length of $2L=4.8$ $\mu$m. The reservoirs are filled with spherical polystyrene particles of diameter (510$\pm$10) nm. We use a laser line trap generated by HOTs to create an attractive potential well that extends from the centre of the channel to 1.7 $\mu$m in the left reservoir (Fig.~\ref{Fig1}(a), (c) and dotted line in Fig.~\ref{Fig1}(d)). After a time interval $T_{\Omega}$, we switch off this laser line and simultaneously switch on a second laser line trap that extends, for a time interval $T_{\Omega}$, from the centre of the channel to 1.7 $\mu$m
in the right reservoir (solid line in Fig.~\ref{Fig1}(d)). In this way, we produce an attractive potential that oscillates at frequency $f=(2T_{\Omega})^{-1}$ between the two channel entrances (Fig.~\ref{Fig1}(a,d) and Video~1). We estimate the extension, depth and shape of the energy wells from the intensity distribution of the line traps generated by HOTs (Fig.~\ref{Fig1}(d))~\cite{Pelton2004,Hayashi2008}. The deduced potentials are validated by measuring the velocity of a particle dragged through the channels by moving the sample stage at a constant speed~\cite{Juniper2012a}. Experiments are performed over a range of oscillation frequencies and particle concentrations $c$ in the reservoirs. 

We record videos of particles undergoing Brownian motion in the channels and reservoirs and extract particle trajectories~\cite{Introduction1996,Dettmer2014b}. We define an attempt as the event for which a particle enters into the channel from either reservoirs and explores it for at least 33 ms, one frame time of the CCD camera that we use~\cite{Pagliara2014e}. Once a particle has entered the channel, it can either go back to the same reservoir, defined as a return event (Fig.~\ref{Fig1}(e)), or translocate through the channel and exit to the opposite reservoir, defined as a translocation event (Fig.~\ref{Fig1}(f)). We determine the attempt rate $J_A$, the translocation rate $J_T$ and the translocation probability $P_T$, defined as ${J_T}/{J_A}$. Average rate values for each oscillation frequency are obtained from at least five experiments of one hour duration each. In order to collect statistically sufficient samples for the translocation time, we use HOTs for automated drag-and-release experiments in which we trap a single particle in one of the two reservoirs and place it in one of the two channel entrances. At $t = 0$ s, we release the particle and simultaneously switch on the oscillating optical potential (sketched in Appendix~\ref{appendix1} and Video~2).

\begin{figure}
\includegraphics[width=\linewidth]{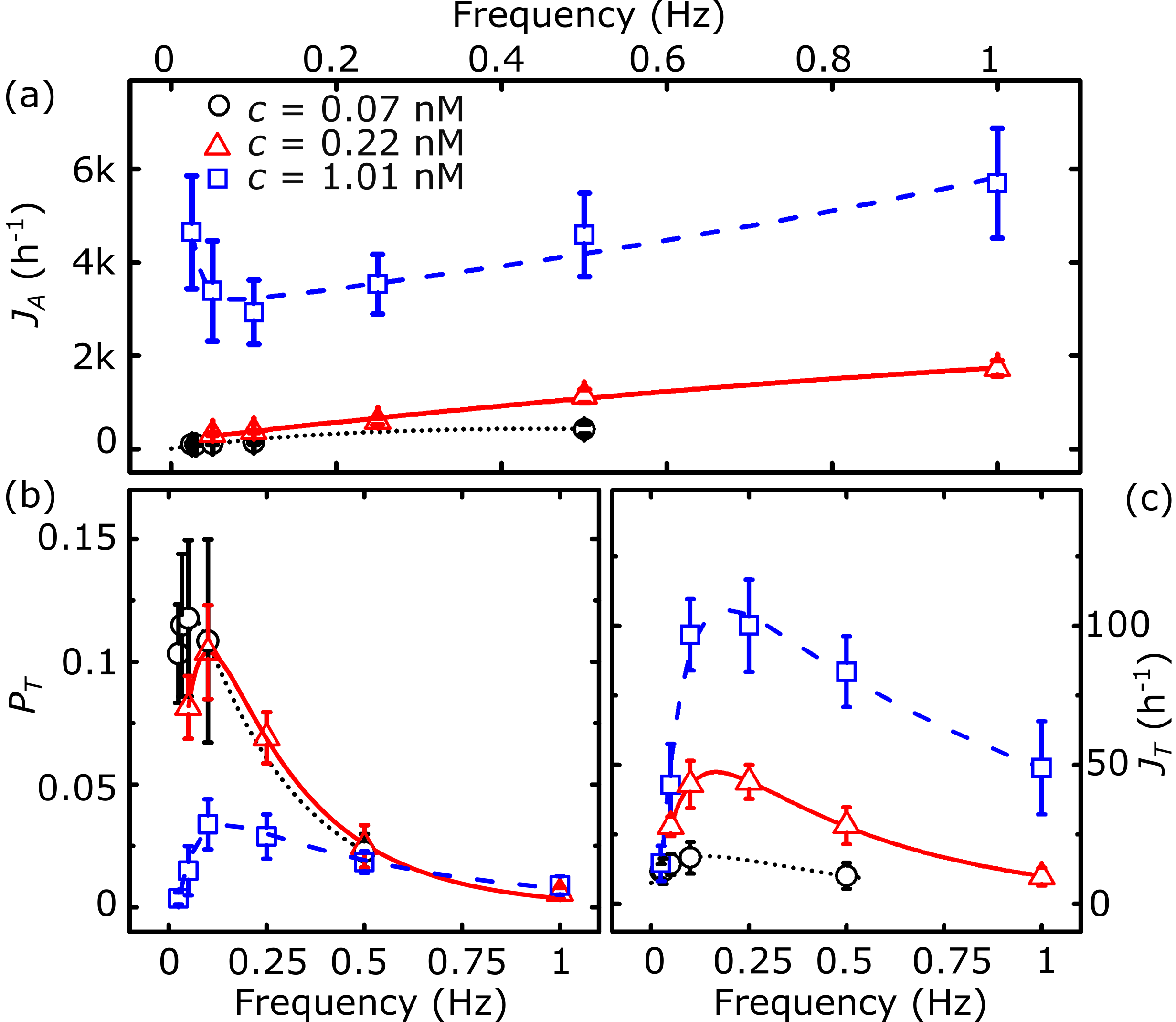}
\caption{\label{Fig2} Dependence of (a) attempt rate $J_A$, (b) translocation probability $P_T$ and (c) translocation rate $J_T$ on the potential oscillation frequency with a particle concentration of 0.07 (circles), 0.22 (triangles) and 1.01 nM (squares) in the reservoirs. Lines are two-term exponential fitting of the data and allow identifying the following peak frequencies: 0.05, 0.10 and 0.15 Hz for the translocation probability and 0.14, 0.16 and 0.19 Hz for the translocation rate at c=0.07, 0.22 and 1.01 nM, respectively (fitting details in Appendix~\ref{appendix3}).} 
\end{figure}

\section{Results and discussion}
\subsection{Dependence of translocation rate and probability on the frequency of the oscillating potential}
For $c = (0.07\pm0.01)$ nM, the attempt rate increases with the oscillation frequency up to $(318\pm30)$ particles (h$^{-1}$) at a frequency of 0.5 Hz (circles in Fig.~\ref{Fig2}(a)). The translocation probability instead sharply decreases down to 0.02 for $f=0.5$ Hz (circles in Fig.~\ref{Fig2}(b)). These two effects cancel each other and as a consequence the translocation rate has a weak dependence on the oscillation frequency (circles in Fig.~\ref{Fig2}(c)).

\begin{figure}
\includegraphics[width=\linewidth]{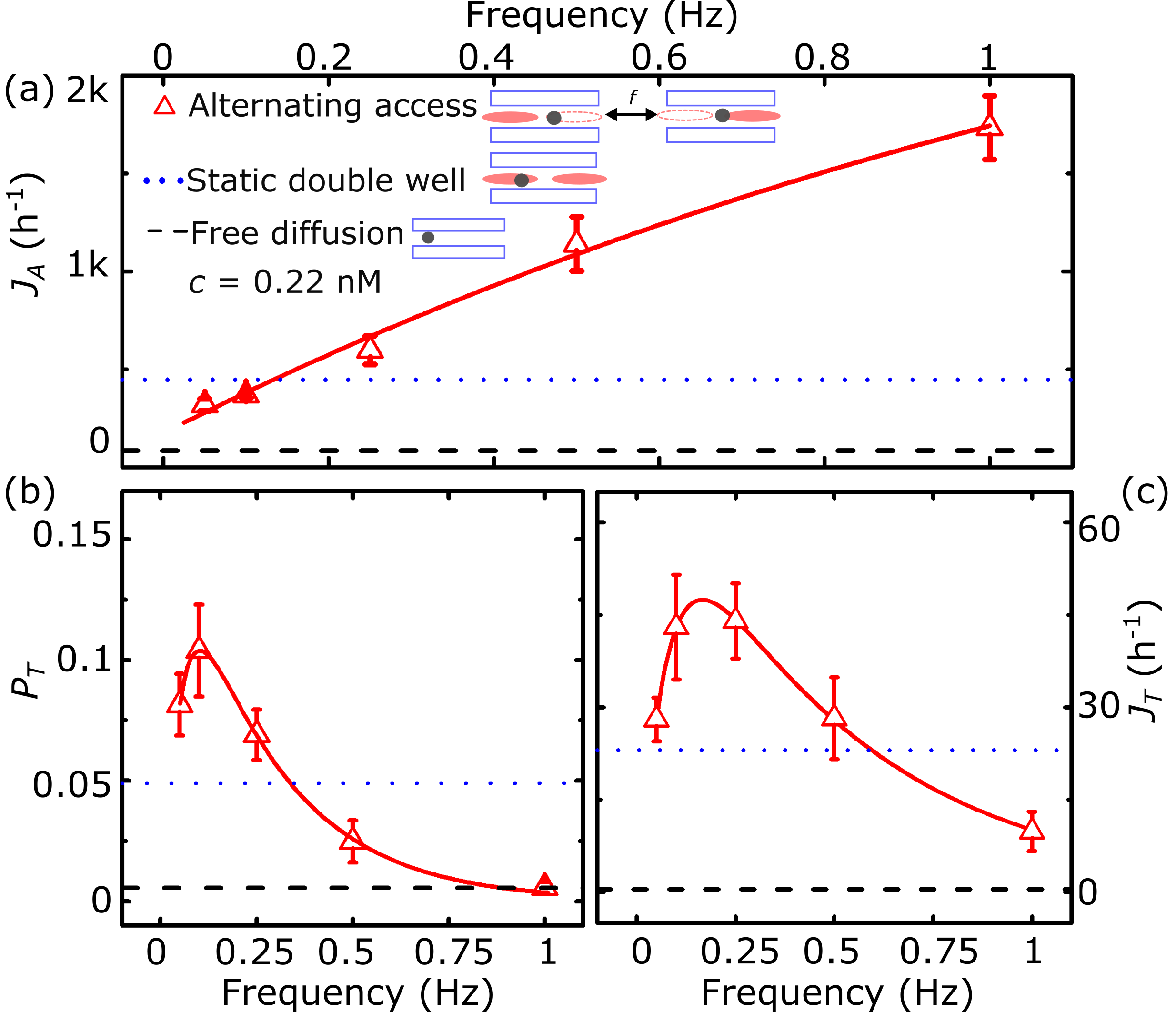}
\caption{\label{Fig3} Comparison of (a) attempt rate $J_A$, (b) translocation probability $P_T$ and (c) translocation rate $J_T$ between the alternating access configuration (triangles), a static potential with two energy wells with the same extension but 42\% smaller depth with respect to the oscillating potential (dotted line) and a channel without optical potential coupled (dashed line). The reduced depth avoids channel jamming (Appendix~\ref{appendix4}) in the presence of a static potential.}
\end{figure}

At higher particle concentrations, the frequency of the oscillating potential strongly affects particle transport across the channel. For $c=(0.22\pm0.06)$ nM, the attempt rate increases with frequency up to a maximum of $(1733\pm163)$ particles (h$^{-1}$) for $f=1$ Hz (triangles in Fig.~\ref{Fig2}(a) and Fig.~\ref{Fig3}(a)). The translocation probability and the translocation rate instead first increase with frequency and peak at oscillation frequencies of 0.1 Hz and 0.16 Hz, respectively (triangles in Fig.~\ref{Fig2}(b,c) and Fig.~\ref{Fig3}(b,c)). 

At even higher particle concentrations $c = (1.01\pm 0.07)$ nM, the attempt rate is not significantly affected by the frequency of the oscillating potential (squares in Fig.~\ref{Fig2}(a)). The translocation  probability and rate first increase with frequency peaking at an optimal oscillation frequency of 0.15 Hz and 0.19 Hz, respectively, and then decrease at higher frequencies (squares in Fig.~\ref{Fig2}(b) and (c)). 

We find that the attempt and translocation rate increase with the concentration of the particles in the reservoirs for all tested oscillation frequencies (Fig.~\ref{Fig2}(a,c)), but interestingly,  this is not the case for the translocation probability (Fig.~\ref{Fig2}(b)). Moreover, for $c = 0.22$ nM the translocation rate at the optimal oscillation frequency $f$ = 0.1 Hz is 102 times higher than the one measured in free diffusion (dashed line in Fig.~\ref{Fig3}(c) and Appendix~\ref{appendix2}) and twice the one measured for a static potential constantly switched on (dotted lines in Fig.~\ref{Fig3}(c)). For $c$ = 1.01 nM at $f=0.1$ Hz, the translocation rate (squares in Fig. 2(c)) is 65 times higher than in free diffusion and 4 times higher than in the presence of the static double well potential.

\subsection{Dependence of channel occupancy on the frequency of the oscillating potential}
\begin{figure}
\includegraphics[width=\linewidth]{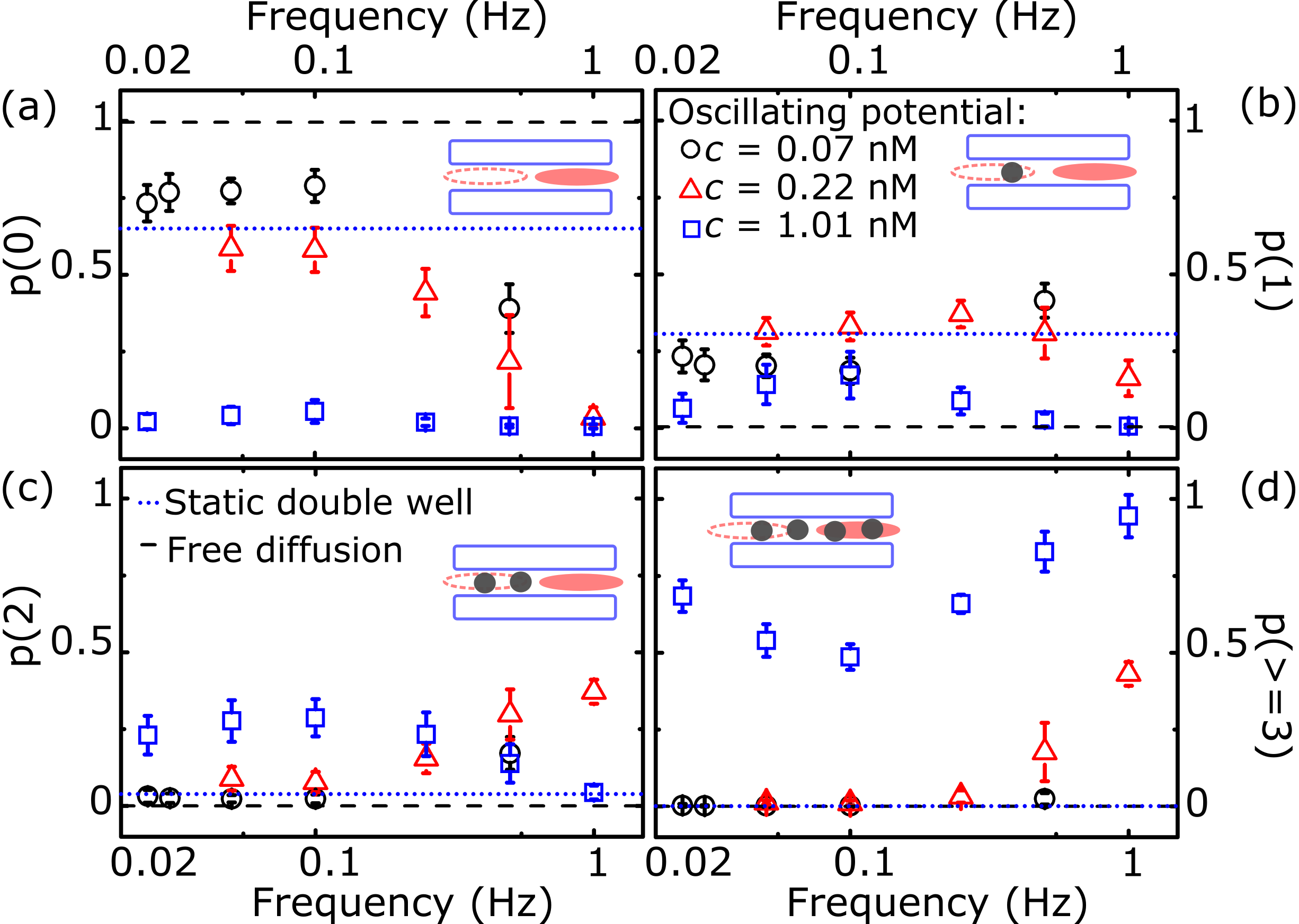}
\caption{\label{Fig4} Dependence of the experimentaly measured probability to find (a) no particle $p(0)$, (b) one particle $p(1)$ , (c) two particles $p(2)$ or (d) more than two colloidal particles $p(\geq3)$ on the oscillation potential frequency and particle concentration. Circles, triangles and squares represent data for a particle concentration of 0.07 nM, 0.22 nM and 1.01 nM, respectively. Dotted and dashed lines represent data for a static potential and  particle free diffusion in the channel, respectively.}
\end{figure}

In order to gain more insight on the presence of an optimal oscillation frequency, we measure the channel occupation probability $p(n)$, which is the probability to simultaneously find $n$ particles in the channel. At high particle concentrations, we measure that the probability to find one particle in the channel $p(1)$ is at a maximum for frequencies close to the optimal oscillation frequency (triangles and squares for $c=0.22$ and 1.07 nM, respectively, in Fig.~\ref{Fig4}(b)). Notably, the channel is predominantly empty at low frequencies (e.g. p(0)=0.58 for $c=0.22$ nM and $f=0.05$ Hz, triangles in Fig.~\ref{Fig4}(a)) whereas at high frequency and concentration the channel is crowded (Fig.~\ref{Fig4}(c,d)), e.g. $p(\geq3)=0.94$ for $c=1.07$ nM and $f=1$ Hz. Therefore, the optimal oscillation frequency is the one that allows for populating the channel without overcrowding it~\cite{Pagliara2014e}. 

\begin{figure}
	\includegraphics[width=\linewidth]{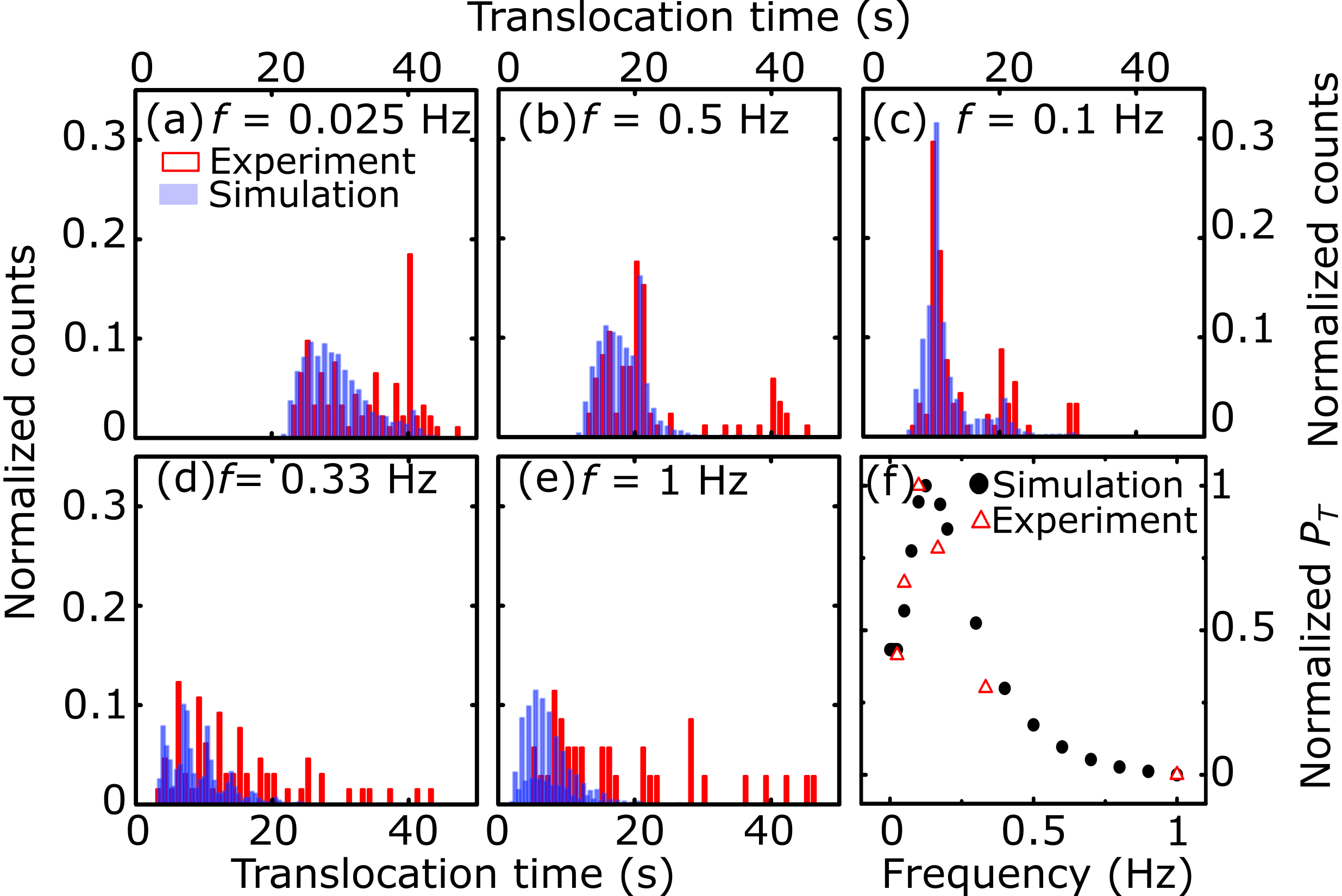}
	\caption{\label{Fig5} (a-e) Histograms reporting the normalized distribution of translocation times measured in drag-and-release experiments at $c=0.01$ nM (red bars) and via Brownian dynamics simulations (blue bars) for different potential oscillation frequencies $f$. (f) Experimental (triangles) and simulated (circles) translocation probability $P_T$ normalized to their maximum values. Experimental $P_T$ is the mean of six sets of 50 independent measurements obtained in drag-and-release experiments.}
\end{figure}

\subsection{Dependence of the channel translocation time on the frequency of the oscillating potential}
We perform drag-and-release experiments to measure the translocation time of a particle across the channel (Video~2). This experiment is repeated at least 300 times for each $f$ and performed at  $c= 0.01$ nM. In free diffusion the translocation time can be calculated as previously reported~\cite{redner2001guide}:
\begin{equation}\label{eq1}
T_{tr}=\frac{2}{3}\frac{L^2}{D_c}
\end{equation}

where $L$ is half the length of the channel and $D_c$ is the particle diffusion coefficient. In the presence of an external potential $U(x)$ in the channel, the translocation time is given by ~\cite{Berezhkovskii2017}:
\begin{equation}\label{eq2}
T_{ext}=\frac{1}{D_c}\frac{\int_{a}^{b}(\int_{a}^{x}\mathrm{e}^{\beta U(y)}\mathrm{d}y)(\int_{x}^{b}\mathrm{e}^{\beta U(y)}\mathrm{d}y)\mathrm{e}^{-\beta U(x)}\mathrm{d}x }{\int_{a}^{b}\mathrm{e}^{\beta U(y)}\mathrm{d}y}
\end{equation}
where $[a,b]$ is the transition length, $\beta=1/(k_BT)$ with $k_B$ and $T$ denoting the Boltzmann constant and absolute temperature.

We measure the distribution of translocation times for each oscillation frequency. We find resonance-like peaks with the first maximum located at 40.5, 20.5, 10.5 and 4.5 s for $f=0.025, 0.05, 0.1$ and 0.33 Hz, respectively (red bins in Fig.~\ref{Fig5}(a-e) and Appendix~\ref{appendix5}). Due to the oscillating nature of our potential, Eq.~(1) and Eq.~(2) can not fully describe our experimental data. However, we performed 1D Brownian dynamics simulations by using the experimentally measured oscillating potential (details in Appendix~\ref{appendix6}) and find translocation time values that favourably compare with the experimental values (Fig.~\ref{Fig5} (a-e)). Moreover, our Brownian dynamics simulations confirm the frequency dependence of the translocation probability with an optimal oscillation frequency of $f = 0.125$ Hz close to the experimentally measured one (black dots and red triangles, respectively, in Fig.~\ref{Fig5}(f)).

\begin{figure}
	\includegraphics[width=\linewidth]{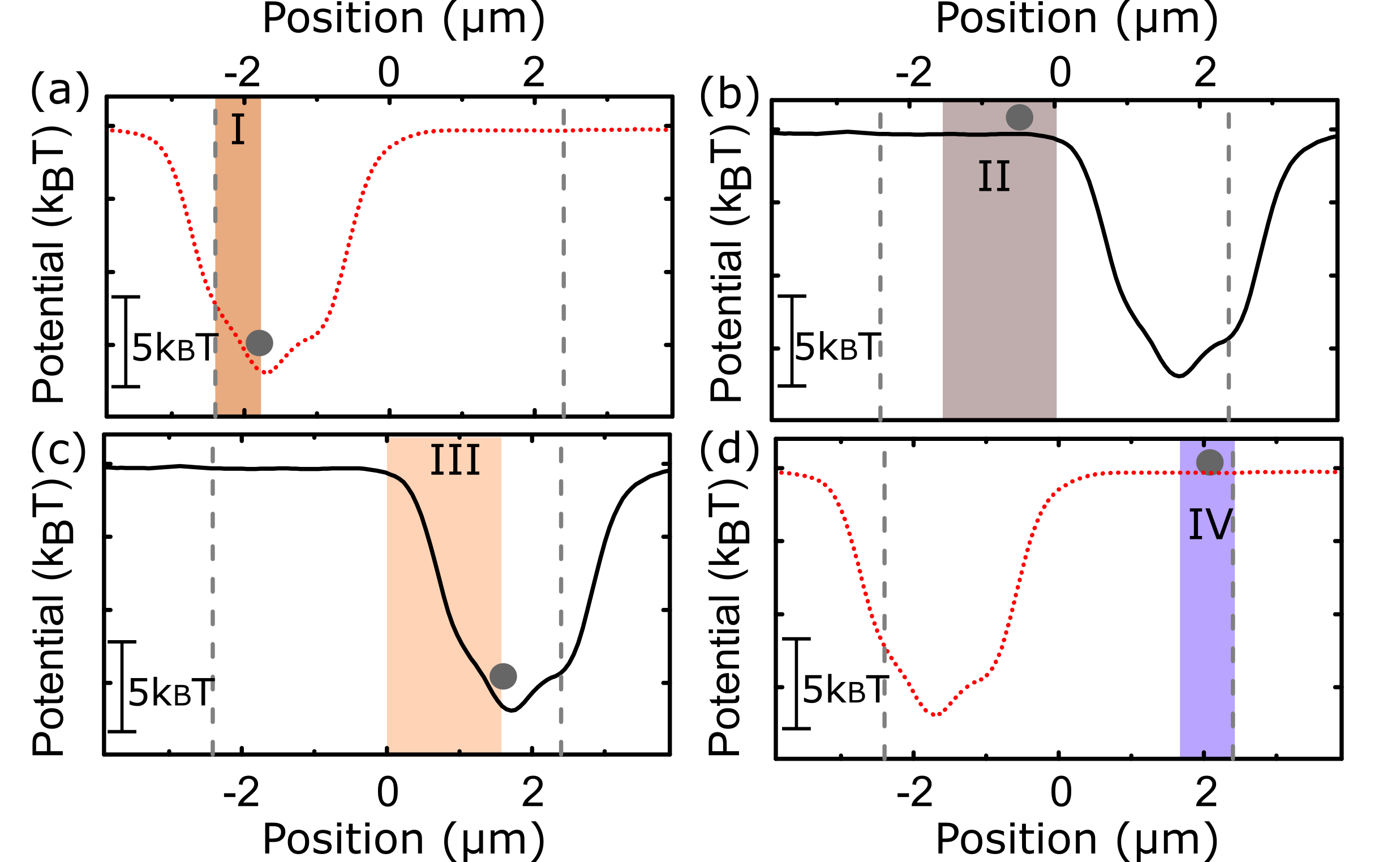}
	\caption{\label{Fig6} Schematics illustrating a representative translocation from left to right in the presence of the oscillating potential. (a) A particle enters the left entrance of the channel when the laser line is on at the left channel entrance and diffuses through region I to the attractive well energy minimum. (b) The left laser line is switched off and simultaneously the right one is switched on, the particle freely diffuses through region II. (c) The particle diffuses through region III into the minimum of the right attractive energy well. (d) The laser line at the right entrance is switched off and the one on the left switched on and the particle diffuses through region IV and out of the channel.}
\end{figure}

\subsection{Optimal oscillation frequency is defined by diffusion between potential wells}
\begin{figure}
	\includegraphics[width=\linewidth]{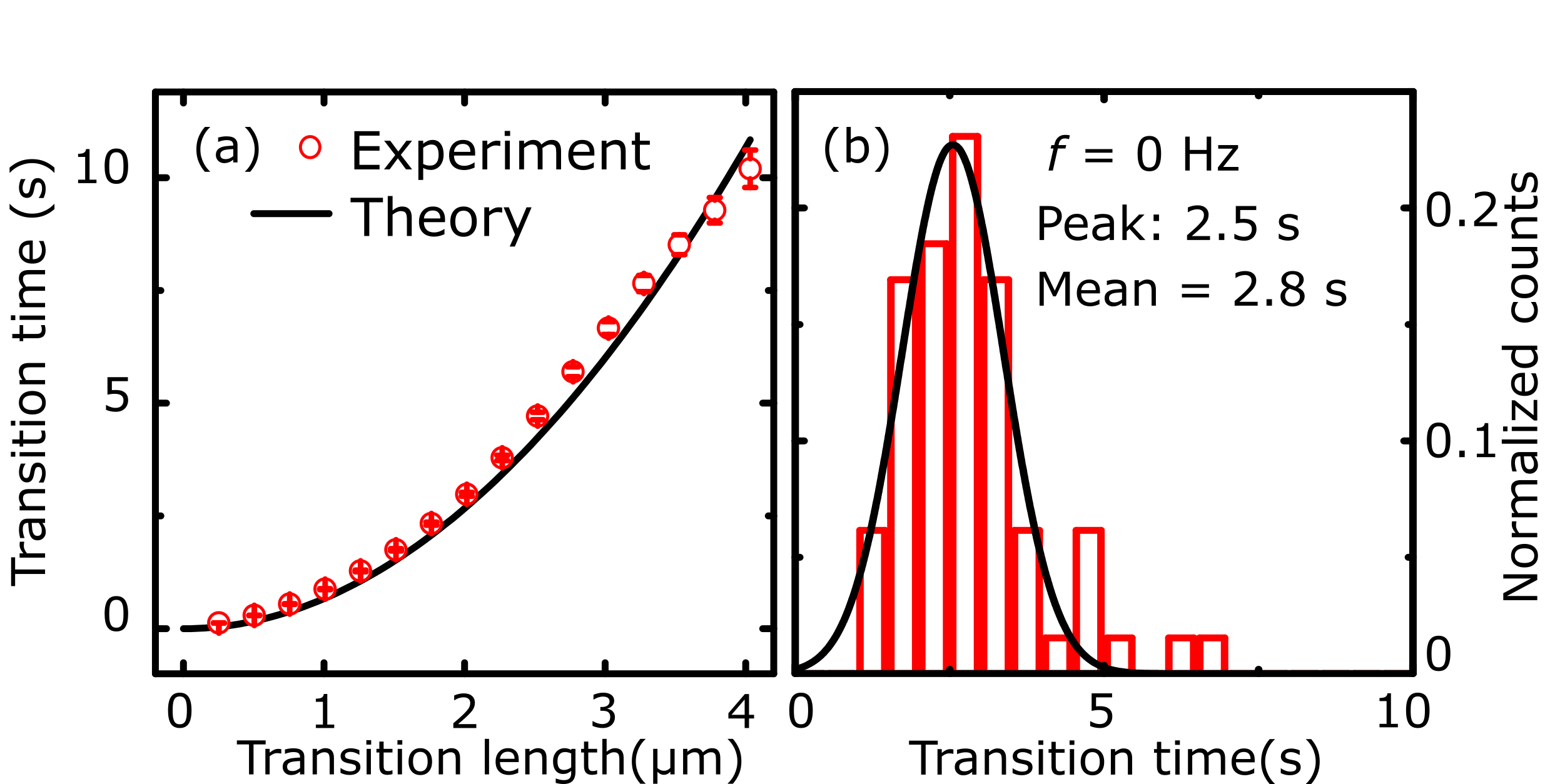}
	\caption{\label{Fig7} (a) Transition times of a freely diffusing particle through portions of the channel of different lengths. The experimental values (circles) are mean and standard errors of the values obtained in 200 experiments. The theoretical values (line) are calculated according to Eq.~1 by using experimentally measured parameters. (b) Distribution of transition times through regions II and III for $f =$ 0 Hz, measured by performing drag and release experiments at $c=0.1$ nM. The peak position is obtained by fitting the data with a Gaussian function.}
\end{figure}

By measuring the transition times across portions of the channel, we provide an intuitive explanation of the optimal oscillation frequency. Firstly, let us consider a representative translocation from the left to the right reservoir. Upon entering the channel, the particle may diffuse to the minimum of the left attractive potential well, while the left laser line is switched on (Region I in Fig.~\ref{Fig6}(a)). The particle is trapped close to this position until, at $T_{\Omega} = (2f)^{-1}$, the left laser line is switched off and the right line is turned on. At this time the particle is free to diffuse either towards the left or right entrance of the channel. In the most efficient scenario in terms of particle transport, the particle travels in free diffusion across region II (Fig.~\ref{Fig6}(b)) and region III where reaches the right-hand side potential minimum when the right laser line is still on (Fig.~\ref{Fig6}(c)). Finally, when this line is switched off the particle is free to diffuse through region IV out of the channel (Fig.~\ref{Fig6}(d)). We perform drag-and-release experiments to evaluate the transition time across each of the four regions above. The particle's first transition time by free diffusion in channel portions of different length is plotted in Fig.~\ref{Fig7}(a). The first transition time through region II and III is $T_{II \& III}=2.8$ s at the oscillation frequency $f = 0$ Hz (Fig.~\ref{Fig7}(b), a particle is released from the HOTs at the left entrance of the channel with the right-hand side potential constantly on). $T_{II}=1.61$ s at $f = 0$ Hz is smaller than the corresponding $T_{tr}=1.95$ s calculated according to Eq.~1 due to the presence of the external potential. $T_{III}=1.02$ s at $f = 0$ Hz is in agreement with the value calculated according to Eq.~2 ($T_{tr} = 1.13$ s) by using the experimentally measured potential. $T_{II \& III}$ is close to $T_{\Omega}=5$ s indicating that the optimal oscillation frequency is the one that matches the transition time through regions II and III. Notably, for $f$ higher than the optimal oscillation frequency, particle's transition through regions II and III is interrupted by the potential oscillation. For $f$ lower than the optimal oscillation frequency, a particle has a higher chance to exit the channel through region I resulting in a return event, although the chance for a particle to be transported through regions II and II is increased. Overall for frequencies different from the optimal frequency,  particle diffusion through regions II and III does not synchronize with the time scale defined by the oscillation frequency. This explains the observed decrease in translocation rate and probability at  $f$ lower and higher than the optimal frequency (Fig.~\ref{Fig2}). 

Optical potentials modulated in time have been extensively employed to direct particle motion~\cite{Lee2006,Juniper2016,Faucheux1995,Leea,Bleil2007,Gorre-Talini1997,MacDonald2003,Jonas2008,Xiao,Ladavac,Simon1992,Babic2004a,Schmitt2006}, including in microfluidic applications~\cite{MacDonald2003}. However, these have yet to be implemented for enhancing particle transport across a quasi 1D microfluidic channel connecting two reservoirs. In this paper we create a modulated optical potential consisting of a laser line that alternates its position between the two entrances of a microfluidic channel. We optimize the oscillation frequency (Fig.~\ref{Fig2}) and explain the physical mechanism underlying the optimal oscillation frequency (Fig.~\ref{Fig4}-\ref{Fig7}). Particle transport in the presence of the modulated optical potential is two orders of magnitudes higher than in free diffusion (Fig.~\ref{Fig3}).

Our experiment indicates that oscillating potentials may be an additional avenue for enhancing transport across synthetic channels or pores. In order to mimic membrane transport in living cells, we are planning to scale our synthetic platform down to the nanoscale~\cite{bell2011dna} where the characteristic diffusion time is closer to the one observed in protein transporter. Furthermore, it is possible to explore asymmetric systems with charged particles only in one of the two reservoirs. Thus our system will be mimicking electrochemical gradients and even exhibit effects like charge polarisation under applied external driving forces. Finally, an avenue with exciting phenomena will involve mixing several types of particles with different sizes and surface charges, exploring competitive effects and thus mimicking the mechanisms of secondary active transporters, where the transport of a substrate is coupled to the transport of a second substrate.

\section{Summary}

We studied the effect of a time dependent potential on particle transport through a microfluidic channel. Inspired by the alternating access mechanism, we coupled an energy well that oscillates between the two entrances of a microfluidic channel. We found that particle transport through the channel can be maximized by optimizing the oscillation potential frequency. Importantly, the optimal oscillation frequency makes the alternating access channel more efficient in terms of transport compared to static channels where particles are either in free diffusion or can simultaneously bind to the ends of the channel. We found that the optimal frequency is the one that allows synchronizing alternating access with particle diffusion across the region of the channel between the two oscillating energy well positions. We anticipate that our findings will stimulate further investigation on mimicking the functioning of membrane protein transporters~\cite{Caspi2008}, on synchronized oscillations~\cite{Doering1992,Schmitt2006,Hayashi2012,Juniper2015} and on the use of modulated potentials for particle control in microfluidics applications~\cite{MacDonald2003}.

\begin{acknowledgments}
We thank Shayan Lameh for proofreading. This work was supported by a Royal Society Research Grant, a Wellcome Trust Strategic Seed Corn Fund and a Start up Grant from the University of Exeter awarded to S.P. U.F.K was funded by an ERC Consolidator Grant (Designerpores 67144). Y.T. was supported by scholarship from Cavendish-NUDT, Lundgren and Pannett Fund, Churchill College. J.G. acknowledges the support of the Winton Programme for the Physics of Sustainability and the European Union's Horizon 2020 research and innovation programme under ETN grant 674979-NANOTRANS. 
\end{acknowledgments}

\appendix
\section{Drag and release experiment}\label{appendix1}
\begin{figure}[H]
	\includegraphics[width=\linewidth]{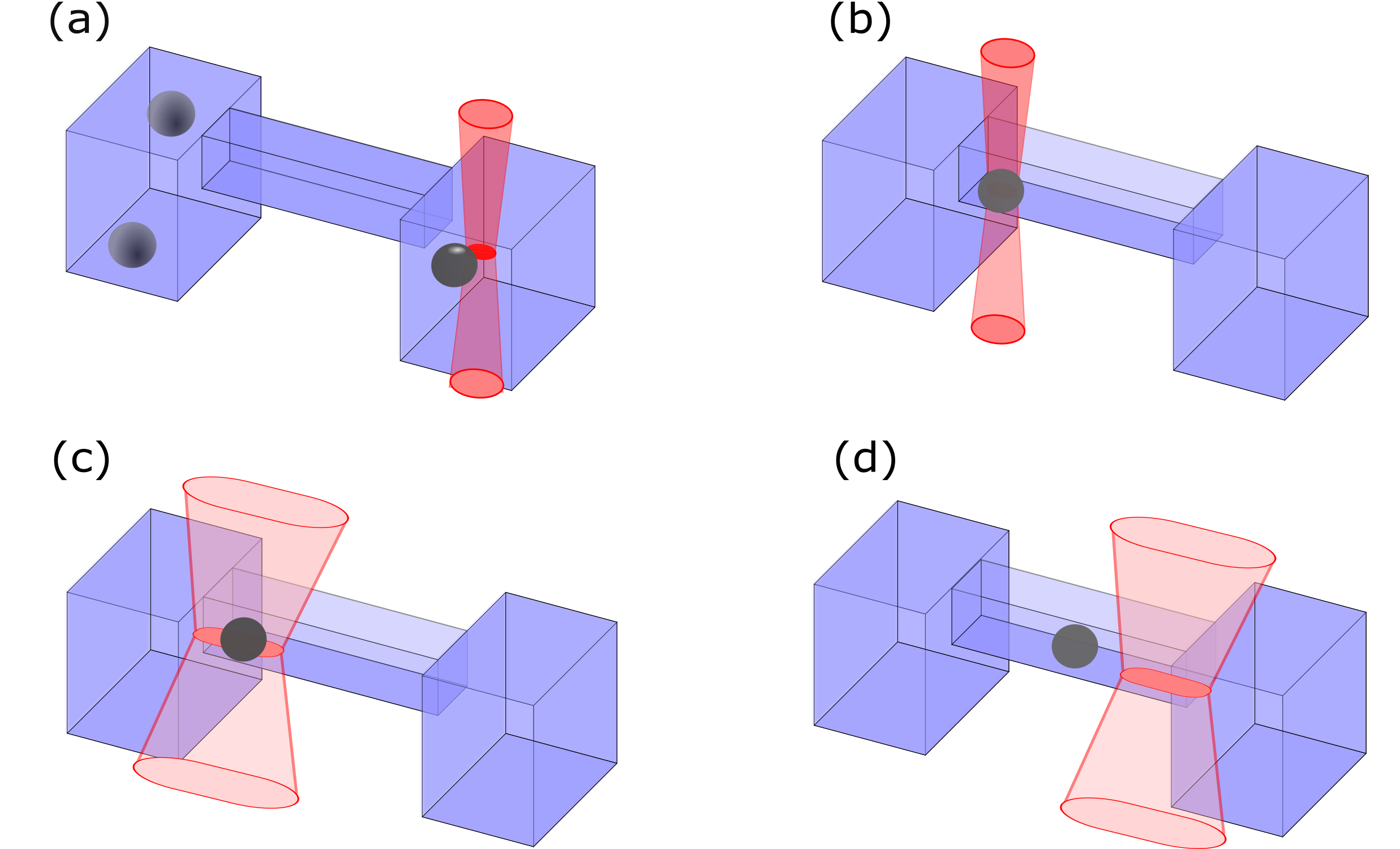}
	\caption{\label{S3} Schematics of the drag-and-release experiment. (a) A particle is trapped in one of the two reservoirs. (b) The particle is dragged and positioned at one of the two channel entrances. (c) The particle is released from the optical trap and simultaneously a laser line is switched on at the same channel entrance. (d) After $T_{\Omega}$ this laser line is switched off and a laser line is switched on at the opposite channel entrance. The laser line position is oscillated at a frequency $f=1/(2*T_{\Omega})$ until the particle leaves the channel. This experiment is repeated at least 300 times for each $f$ and performed at particle concentration $c= 0.01$ nM in the reservoirs.}
\end{figure}

\section{Transport of particles through the channel in free diffusion}\label{appendix2}
\begin{figure}[H]
	\includegraphics[width=\linewidth]{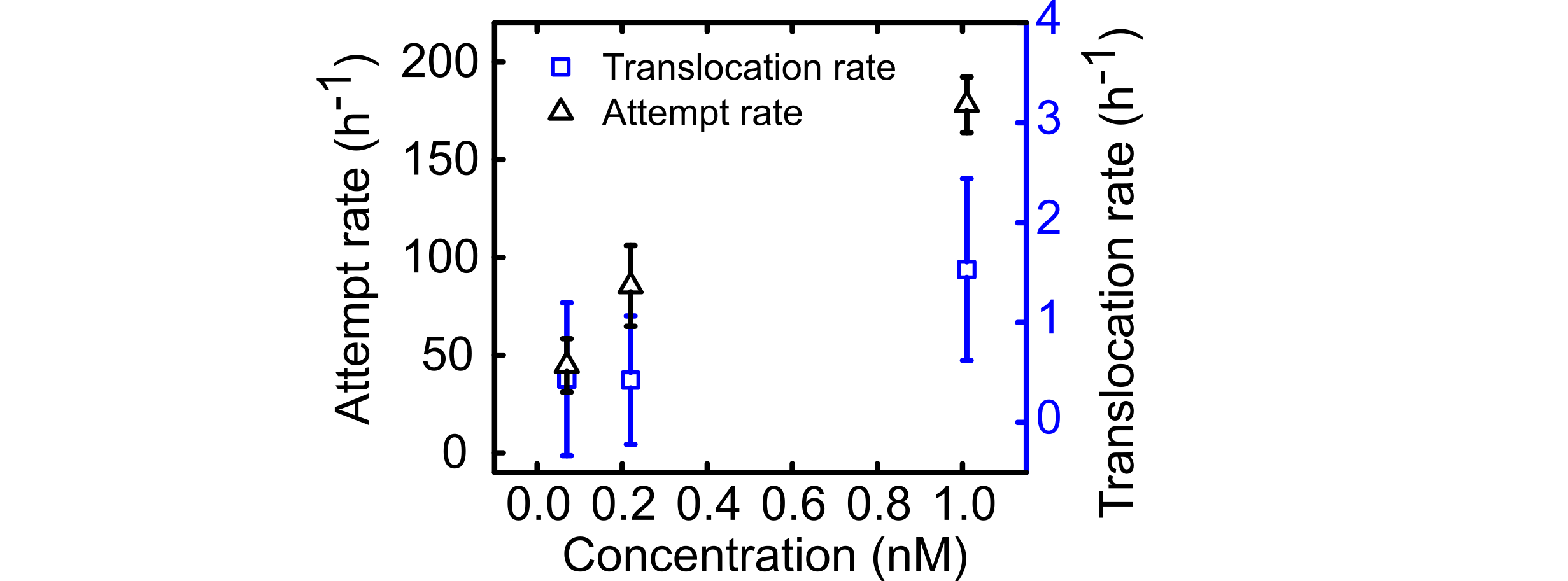}
	\caption{Dependence of the attempt (triangles) and translocation rate (squares) with respect to particle concentration for particles in free diffusion through the microfluidic channel described in the main text. Data and error bars are the mean and standard deviation of the values measured in five different one hour long experiments.}
\end{figure}

\section{Fitting attempt and translocation rate}\label{appendix3}

Attempt rate, translocation rate and translocation probability in Fig.~2 and 3 are fitted by a two-term exponential model $f(x) = a*\exp(b*x) + c*\exp(d*x)$ via the nonlinear least-squares method, where $a,b,c,d$ are the fitting parameters. The values for these parameters estimated by the fitting are reported in Tables I-III.
\begin{table}[H]
	\caption{Fitting of attempt rate data}
	\begin{ruledtabular}
		\begin{tabular}{ccccccc}
			$c$ (nM)& a& b& c&  d 
			\\
			\hline
			0.07& 9853436.5& -2.1& -9853424.6& -2.1\\
			0.22& 83037687.0& -0.3& -83037511.7& -0.3\\
			1.01& 3013.6& 0.65& 10014.3& -73.7\\
		\end{tabular}
	\end{ruledtabular}
\end{table}

\begin{table}[H]
	\caption{Fitting of translocation rate data}
	\begin{ruledtabular}
		\begin{tabular}{ccccccc}
			$c$ (nM)& a& b& c&  d 
			\\
			\hline
			0.07& 26.7& -1.9& -19.0& -12.7\\
			0.22& 77.4& -2.0& -86.7& -14.5\\
			1.01& 137.6& -1.0& -187.9& -17.0\\
		\end{tabular}
	\end{ruledtabular}
\end{table}

\begin{table}[H]
	\caption{Fitting of translocation probability data}
	\begin{ruledtabular}
		\begin{tabular}{ccccccc}
			$c$ (nM)& a& b& c&  d 
			\\
			\hline
			0.07& 0.2& -3.9& -0.1& -45.0\\
			0.22& 0.2& -3.9& -0.2& -23.9\\
			1.01& 0.05& -1.8& -0.1& -20.7\\
		\end{tabular}
	\end{ruledtabular}
\end{table}

\section{Channel jamming with static double well potential}\label{appendix4}

\begin{figure}[H]
	\centering
	\includegraphics[scale=1]{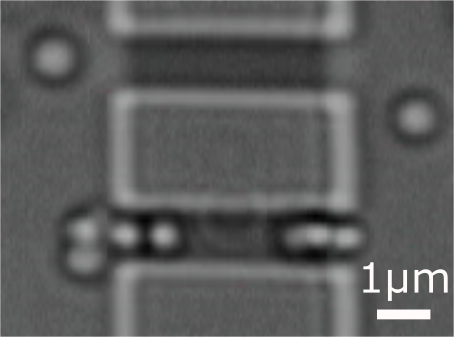}
	\caption{\label{S2} Bright field image of 510 nm polystyrene particles jamming the entrances of the channel permeated with static energy wells as deep as the potential wells used for the oscillation. For this reason in the experiments reported in Fig.~3 and Fig.~4, for the static potential we used a depth $42\%$ smaller with respect to the well depth used for the oscillating potential.}
\end{figure}

\section{Transition and translocation times}\label{appendix5}
For a particle in free diffusion in the channel the translocation time calculated according to Eq.~1 using $L=4.8/2$ $\mu$m and $D_c$ = 0.25 $\mu m^2/s$~\cite{Dettmer2014b} , is 15.36 s. This is also in agreement with the value obtained via Brownian simulation (14.99 s).

\begin{figure}[H]
	\includegraphics[width=\linewidth]{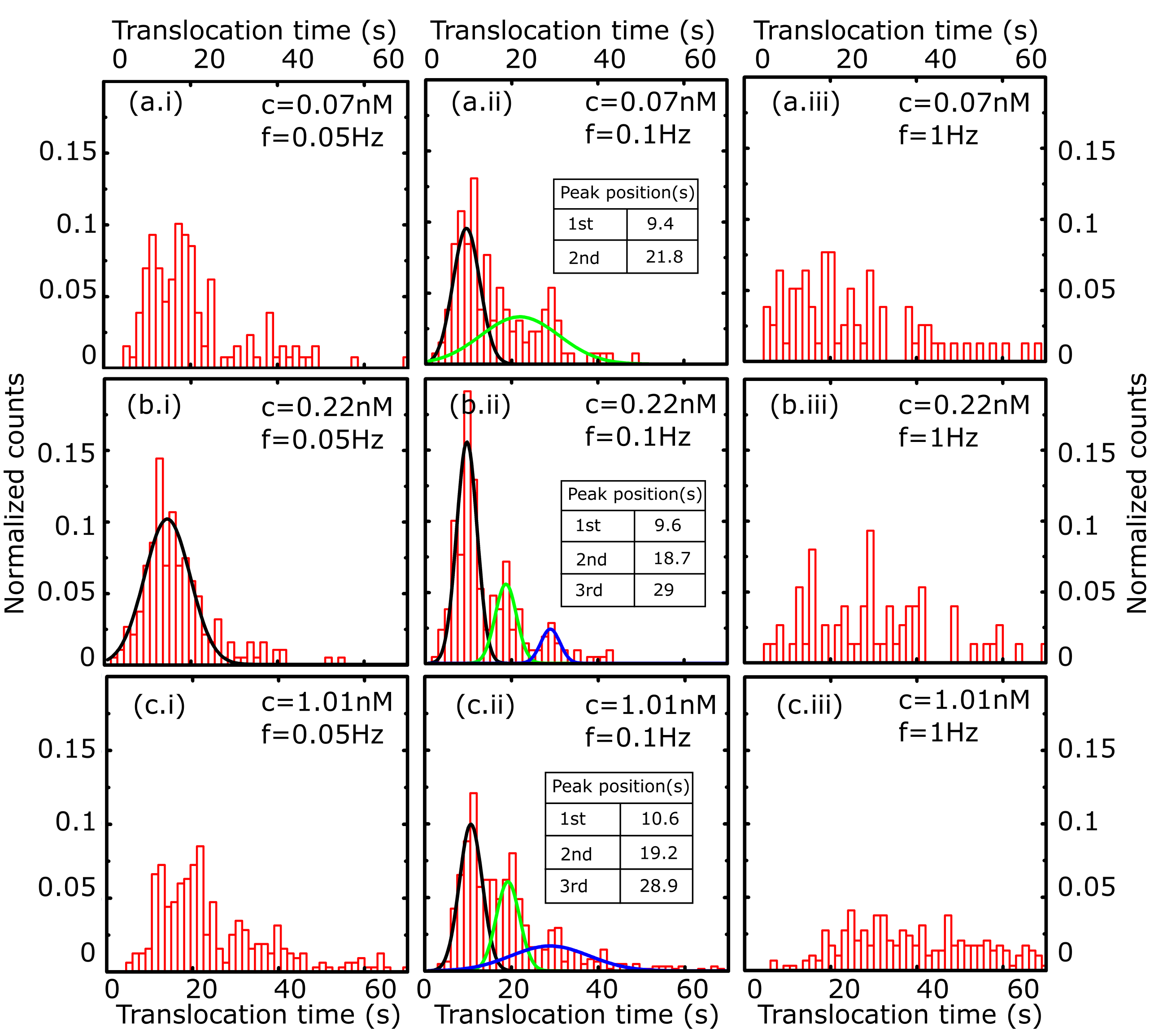}
	\caption{\label{S7} Histograms reporting the measured distributions of translocation times for $f = 0.05$ Hz, 0.1 Hz and 1 Hz (from left to right) and c = 0.07 nM, 0.22 nM and 1.01 nM (from top to bottom). The distributions are fitted with Gaussian functions (solid lines) allowing for the extrapolation of harmonic peaks. The insets report the harmonic peak positions obtained by fitting Gaussian functions.  }
\end{figure}

\section{Brownian dynamics simulation}\label{appendix6}
We carried out Brownian dynamics simulations in a 1D channel with length $L$. 

Particle trajectories start at $-L/2$ and are terminated at their first contact with each of the perfect absorbing boundaries set at $-L/2$ and $L/2$. In simulations we track the fraction of particles that end at $L/2$ as well as their transit time defined as the time that a particle takes to reach $L/2$ for the first time. 

The actual particle's position, $x_{n+1}$, is given by $x_{n+1}= x_n + x_{ran} + \beta D_c F \Delta t$, where $x_n$ is the previous position, $x_{ran}$ is a pseudo random number generated with a Gaussian distribution with average position displacement $\mu = 0$ and standard deviation $\sigma = \sqrt{2 D_c \Delta t}$, and Force $F$ was derived from the potential depicted in Fig.~1(d). When running simulations we set $\Delta t = 1\times 10^{-4}$ s and we average over 100 millions random walkers. 

\newpage
\bibliography{Oscillating_transport_cite}
\newpage
\end{document}